\documentclass[12pt,preprint]{aastex}

\def\lessim{\mathrel{\hbox{\rlap{\hbox{\lower4pt\hbox{$\sim$}}}\hbox{$<$}}}}
\def\grtsim{\mathrel{\hbox{\rlap{\hbox{\lower4pt\hbox{$\sim$}}}\hbox{$>$}}}}

\shorttitle{Photometry of Swift~J2319.4+2619}
\shortauthors{Shafter et~al.}

\begin{document}


\title{Time-Resolved Photometry of the Optical Counterpart of\\
Swift~J2319.4+2619}


\author{A. W. Shafter, J. R. A. Davenport, T. G\"uth, S. Kattner, E. Marin,\\and N. Sreenivasamurthy}
\affil{Department of Astronomy and Mount Laguna Observatory\\
     San Diego State University\\
    San Diego, CA 92182}
\email{aws@nova.sdsu.edu,jdavenpo@sciences.sdsu.edu,tgueth@sciences.sdsu.edu,\\skattner@sciences.sdsu.edu,emarin@sciences.sdsu.edu,nivedita\_star@yahoo.co.in}




\begin{abstract}

Time-resolved CCD photometry is presented of the $V\sim17$ optical
counterpart of the newly-discovered, hard-X-ray-emitting
polar Swift~J2619.4+2619.
A total of $\sim$20~hr of data obtained over five nights
in various bandpasses ($B, V, R,$ and $I$)
reveals a strong quasi-sinusoidal
modulation in the light curve at a best-fitting
period of 0.1254~d (3.01~hr),
which we associate with the orbital period of the system
(one-day aliases of this period at 0.1114~d and 0.1435~d are
considered, but appear
to be ruled out by our analysis). The amplitude of the modulation
increases with wavelength from $\sim$0.8~mag in $B$ to $\sim$1.1~mag in $R$
and $I$. The increase in amplitude with wavelength is typical of
polar systems where
the modulated radiation comes from cyclotron emission. The combination
of the relatively long orbital period and the emission of hard X-rays
suggest that Swift~J2619.4+2619 may be a good candidate for an
asynchronous polar system.

\end{abstract}





\section{Introduction}

The magnetic cataclysmic variables (mCVs)
form a subclass of the cataclysmic variables stars consisting
of a low-mass, late-type dwarf that fills its Roche lobe and transfers
mass to a magnetic white dwarf companion (e.g. see
Warner 1995, Wickramasinghe \& Ferrario 2000). In the most strongly
magnetic systems, the polars
(a.k.a. AM~Her stars),
the white dwarf's magnetic field is sufficiently strong to lock it into
synchronous rotation with the orbit and to inhibit the formation of
an accretion disk, forcing the transferred gas
to accrete onto one (or sometimes both) of the white dwarf's
magnetic poles. In the intermediate
polars (IPs, a.k.a. DQ~Her stars), the magnetic field
strength is insufficient
to phase lock the white dwarf, and accretion occurs onto both magnetic
poles via a truncated accretion disk.
In recent years, a third type of mCV has been recognized,
the asynchronous polars, where the white dwarf rotates slightly
($\sim1-2$\%) out of synchronism with the orbit (Campbell \& Schwope 1999). 
In all mCVs, a high-temperature ($T\grtsim10^8$~K)
shock is formed at the base of the accretion column, resulting in the
emission of X-rays through bremsstrahlung radiation.
At the relatively high magnetic field
strengths typical of polar systems ($B\grtsim10$~MG), a significant
component of the cooling in the post-shock gas also occurs via 
polarized optical and infrared cyclotron radiation, which tends
to lower the temperature of the post-schock gas and soften the X-ray
spectrum (Lamb \& Masters 1979). Since the cyclotron radiation is
beamed preferentially in directions perpendicular to the field lines,
the changing aspect of the accretion column
results in a significant modulation in the optical and infrared
light curves at the orbital period of the binary.

A series of X-ray and optical spectroscopic
observations reported in Mukai et al. (2007)
have shown that the recently identified
Swift/BAT source, Swift~J2319.4+2619, is likely to be
a rare, hard-X-ray-emitting polar.
The object was detected initially as a
Swift/BAT source with a $15-100$~keV spectrum that could be fitted with
either a power law of photon index 2.7,
or a bremsstrahlung spectrum (kT=19 keV),
with a $15-50$~keV flux of $1.05 \times 10^{-11}$~ergs~cm$^{-2}$~s$^{-1}$.
Later, pointed Swift/XRT observations obtained in May and June 2007 enabled
Mukai et al. to identify Swift~J2319.4+2619 both with the
ROSAT All-sky Survey source 1RXS~J231930.9+261525, and with a
probable optical counterpart, USNO~B1~1162-0585089.
Follow-up spectroscopic observations
then revealed strong Balmer and He~II emission
lines typical of polars.
In this paper we present time-resolved, multi-color, CCD photometry
of the optical counterpart of Swift~J2319.4+2619, first reported
in Shafter et al. (2007), which has enabled us to
determine the likely orbital period of the system.
We conclude by discussing the possibility that future observations may
reveal Swift~J2319.4+2619 to be a member of the class of asynchronous polars.
 
\section{Observations}

A finding chart for the optical counterpart to
Swift~J2319.4+2619 is shown in Figure~1,
with the coordinates taken from Mukai et al. (2007).
Observations
were carried out during five nights in December 2007 using
the Mount Laguna Observatory 1~m reflector.
Each night, a series of one minute
exposures were taken through either a Johnson-Cousins
$B, V, R, {\rm or}~I$ filter (see Bessel 1990),
and imaged on a Loral $2048^2$ CCD.
To decrease the read-out time between exposures, only a $600\times600$
subsection of the
full array was read out. The subsection was chosen to include
Swift~J2319.4+2619 and
several relatively bright nearby field stars to be used for
differential photometry.
A summary of observations is presented in Table~1.

The data were processed in a standard fashion (bias subtraction and
flat-fielding) using IRAF.\footnote{
IRAF (Image Reduction and Analysis Facility) is distributed by the
National Optical Astronomy Observatories, which are operated by AURA, Inc.,
under cooperative agreement with the National Science Foundation.}
The individual images were subsequently
aligned to a common coordinate system and
instrumental magnitudes for Swift~J2319.4+2619
and a nearby comparison star were then determined using the
{\it IRAF\/} APPHOT package.
Variations in atmospheric transparency were removed to first order by dividing
the flux of Swift~J2319.4+2619 by that of a nearby comparison star
located approximately
$2.5'$~W and $1'$~S of the variable (star ``C" in Fig. 1).
The differential light curves were then placed on an
absolute scale by calibrating the comparison star against
standard stars from Landolt (1992).
For the comparison star, C, we find
$V=13.26\pm0.10$, $B-V=0.98\pm0.10$, $V-R=0.49\pm0.10$, and $V-I=1.06\pm0.10$.
The final calibrated light curves of Swift~J2319.4+2619 are
displayed in Figure~2.

\section{The Orbital Period}

In polar systems a significant component of the optical radiation
comes from cyclotron emission radiated from the accretion column.
Since the white dwarf's rotation (and hence its magnetic axis) is
locked into synchronism with the orbit,
polar systems show strong modulations in their
optical radiation at the orbital period of the binary.
The nature of the modulation depends largely on the system
geometry; specifically, the orientation of the magnetic axis relative to the
spin axis, and the orientation of the orbital plane to
our line of sight.
In cases where the
accreting column is visible throughout the orbit, the light
curve is quasi-sinusoidal, with the observed modulation arising
from the combined effects of cyclotron beaming and the changing
orientation of the magnetic field axis to our line of sight.
Alternatively, if the accreting magnetic pole
passes behind the limb of the white dwarf during part of the
orbit, the light curve will be characterized by flat intervals
(during self eclipse of the accretion column) followed by
broad ``humps" when the column is visible.
The former systems are often referred to as ``one-pole" systems
because the accreting pole is the only pole ever visible, with
the latter systems, which have
the non-accreting pole visible during the part of
the orbit when the accreting pole is hidden,
referred to as ``two-pole" systems (Cropper 1990).

The photometry of Swift~J2319.4+2619 shown in Figure~2
clearly reveals a quasi-sinusoidal modulation
with a time scale of $\sim$3~hr, indicating that
the object is a one-pole system. As expected for cyclotron emission
in polars, the amplitude of the
modulation increases with wavelength, in this case with values ranging from
0.8~mag in $B$ to 1.1~mag in $R$ and $I$.
We determined the period of the modulation using two approaches.
Initially, we estimated times of maximum light from
the point of intersection of straight line segments fitted by eye to the
ascending and descending portions of the hump profiles.
During the five nights of observation, we observed a total of eight
maxima. The resulting timings, which we estimate to be accurate to $\pm5$~min,
are given in Table~2. A linear least-squares fit of these timings
yields the following ephemeris for the times of
maximum light in Swift~J2319.4+2619:
\begin{equation}
T_{\mathrm{max}} = {\rm HJD}~2,454,437.618(2)+0.12544(3)~E.
\end{equation}
Residuals of the measured timings with respect to eqn~(1)
are also included in Table~2.

Next, to make a more thorough period search,
including possible alias periods,
we analyzed the entire data set
by computing a Scargle (1982) periodogram.
Given that the light curves were obtained in different colors
with differing amplitudes and zero-points,
we converted all
light curves to relative flux with maxima normalized to unity
prior to computing the periodogram.
The results of this analysis is shown in Figure~3.
The 0.1254~d period determined above is clearly favored, however,
it is not obvious that the
one cycle per day aliases of this period
(0.1114~d and 0.1435~d) can be ruled out definitively.

In order to explore the viability of these alias periods,
we have performed two additional exercises. First, we have reproduced
in Figure~4 the light curves from the first three nights of
observation, where we covered two maxima, showing the times
of maxima expected for the best-fitting period and its one-day
aliases. We have chosen to anchor the timings to the second
maxima in each case because these maxima are the best defined
(most symmetrical). Clearly, the shorter alias period ($P=0.1114$~d,
dotted line) appears to be too short to match the light curves,
particularly in the case of the $R$-band light curve.
On the other hand, the longer alias period ($P=0.1435$~d,
dashed line) clearly appears to be too long in all cases.

As a final test, we have phased the light curves with each of the
three periods, and plotted the normalized and
phased light curves in Figure~5. The 0.1254~d period clearly produces
the phased light curve with the least scatter, indicating that it
provides the best fit to the data. Taken together, the analyses
presented above strongly favor the 0.1254~d period, and we adopt
this as the likely orbital period of Swift~J2319.4+2619.

\section{Discussion}

Swift~J2319.4+2619 is somewhat unusual (for a polar) given the strength of
its {\it hard} X-ray emission.
It has long been recognized that among mCVs,
polars have relatively soft X-ray spectra compared with the
IP systems (Kuijpers \& Pringle 1982; Lamb 1985).
The ratio of hard-to-soft X-ray emission in a mCV
depends strongly on the magnetic moment of the white dwarf.
White dwarfs with high magnetic moments (high fields and low masses)
are expected to have strong, soft X-ray emission, with low-field,
high-mass systems expected to radiate significant hard X-ray emission
(e.g. see Lamb \& Masters 1979, Cropper et al. 1998).
The dominant post-shock cooling mechanism in high-field systems,
such as polars, is cyclotron emission, which tends to decrease the
post-shock temperature and suppress hard X-ray emission, as would
a lower mass white dwarf.
On the other hand, the higher temperatures required to
produce the hard X-ray emission seen in IPs can be produced
in a lower-field system, which would be expected to cool
primarily through bremsstrahlung radiation, or in a
system with a high-mass white dwarf, or both.
In a recent study of cataclysmic variables discovered in the
INTEGRAL/IBIS survey, Barlow et al. (2006) found that IP systems
(and the asynchronous polar systems)
were significantly more likely to be detected in the hard 20-100~keV
energy band than were polar systems. In the sample of 19 cataclysmic
variables discovered
by IBIS, 14 are known or suspected magnetic systems. Of these,
V834~Cen is the only polar. Among the rest,
11 are IPs, two are asynchronous polars, and one is a bright (and nearby)
dwarf nova (SS Cyg).

The majority (13)
of the 19 hard X-ray bright systems have known orbital periods,
and all but one, the polar V834 Cen,
have periods in excess of the well-known $2-3$~hr gap
in the orbital period distribution of cataclysmic variables.
Although it appears that the gap is less well defined for
mCVs, it remains true
that the IPs typically have orbital periods above
the gap, whereas polar systems are mostly concentrated
at orbital periods below the gap (e.g. Webbink \& Wickramasinghe 2002).
The generally accepted explanation
for this segregation is that the smaller binary dimensions and
lower mass accretion rates characteristic of the short period systems
make it easier for white dwarfs with a given magnetic moment
to become locked in synchronism with the orbit (e.g. King et al. 1985).
Patterson (1994) has estimated the minimum field strength necessary to
phase lock a white dwarf of a given mass, $M_{wd}$, in a system
with orbital period, $P_{orb}$, as
\begin{equation}
B_7\grtsim1.5M_{wd}^{3.1}P_2^{2.8},
\end{equation}
where $B_7$ is the magnetic field strength in units of $10^7$~G,
$M_{wd}$ is the white dwarf mass in solar units, and $P_2=P_{orb}/2$~hr.

The orbital period of Swift~J2319.4+2619, as determined
from the observed photometric modulation, places the system just
above the $2-3$~hr period gap. The combination of the relatively
long orbital period (for a polar), coupled with
the strong hard X-ray emission, suggests that it may be difficult to
phase-lock the white dwarf in this system. In particular, as noted earlier
the hard X-ray emission suggests either a low field strength, or high
white dwarf mass, or both. However,
as can be seen from eqn~(2), these requirements
make synchronism increasingly difficult to achieve at long orbital periods.

\subsection{Is Swift~J2319.4+2619 an Unrecognized Asynchronous Polar?}

Taken together, the observed properties of
Swift~J2319.4+2619 raise the possibility
that the system may be a member of the (rare?) class of asynchronous polars.
In such systems, of which there are currently only four known
(BY~Cam, V1500~Cyg, V1432~Aql, CD~Ind),
the white dwarf spin period and the orbital
period differ by $\sim1-2$\% (Patterson et al. 1995,
Campbell \& Schowpe 1999). The cause of the slight asynchronism
in these systems is unknown, but may be caused by recent
(recorded or unrecorded) nova
eruptions (V1500~Cyg is a known nova, Nova Cyg 1975), or, as suggested
by Patterson et al. (1995), by the difficulty
in synchronizing white dwarfs at longer orbital periods
(among the four known asynchronous polars, all but CD~Ind have
orbital periods in excess of 3.3~hr). The latter explanation
would seem to be relevant to BY~Cam and
V1432~Aql, where the orbital periods ($\sim$3.36~hr) and estimated
white dwarf masses ($\sim$1~M$_{\odot}$, Ramsay 2000) suggest via eqn~(2)
that $B\grtsim60$~MG would be required to synchronize the white
dwarfs in these systems. Although the field strength in V1432~Aql
is unknown, in BY~Cam it is estimated to be $\sim$41~MG (Cropper et al. 1989),
which would appear insufficient to assure synchronism of the white dwarf.

It is possible that
asynchronous polar systems may be considerably more common than
is implied by the small number of systems currently known.
In order to classify a mCV
as an asynchronous polar, it is clearly necessary
to establish the orbital period independently of any
photometric (or polarimetric) modulation tied to
the white dwarf rotation. The orbital periods in polars are
difficult to measure directly unless the system is eclipsing.
For non-eclipsing systems, such as Swift~J2319.4+2619,
the orbital period can be measured
directly only if spectral features from the secondary star are visible,
and a radial velocity curve can be determined. Future
observations of Swift~J2319.4+2619, particularly if the
system enters a low-state when the relative contribution
of the secondary star's light to the overall system luminosity
is increased, should focus
on the detection of spectral features from the secondary star.
For an orbital period of 3~hr, the secondary star is expected to
have a spectral type of $\sim$M$4-5$ (Smith and Dhillon 1998),
and would be most easily detected in the near infrared, for example
through the Na~I doublet at 8183, 8194\AA.

\section{Conclusions}

We have presented five nights of multicolor CCD photometry of
the optical counterpart of Swift~J2319.4+2619, which has
been recently identified
as a hard-X-ray-emitting polar by Mukai et al. (2007).
Our principal conclusions are as follows:

1) A strong, quasi-sinusoidal photometric modulation is seen in the
light curve with a best-fitting period of 0.1254~d (3.01~hr),
which we identify as the likely orbital period
of the system. A detailed analysis of the data appears to
rule out the one-day aliases of this period at P=0.1114~d (2.67~hr) and
P=0.1435~d (3.44~hr) as viable orbital period candidates.

2) The sinusoidal
nature of the modulation suggests that Swift~J2319.4+2619 is a
``one-pole" accretor,
however a determination of the detailed accretion geometry
and magnetic field strength
will require future time-resolved polarimetric observations.

3) The amplitude of the modulation ranges from $\sim$0.8~mag in $B$
to $\sim$1.1~mag in the $R$ and $I$ bands.
The slight increase in the amplitude of the modulation
with wavelength is characteristic of the cyclotron emission produced
at typical polar field strengths, providing additional support for
the identification of Swift~J2319.4+2619 as a polar system.

4) The relatively long orbital period, coupled with
strong hard X-ray emission observed by Mukai et al. (2007)
suggests that Swift~J2319.4+2619 may be an asynchronous polar.
In order to determine if
the white dwarf rotates synchronously with the orbit,
future observations should focus on detecting radial velocity
variations of the secondary star, which will allow
an independent determination of the orbital period of the system.



\acknowledgments
We would like to thank the referee for constructive comments regarding
X-ray emission in mCVs.
This research was supported in part by NSF grant AST-0607682.

\clearpage

\begin{deluxetable}{ccccc}
\tablenum{1}
\tablewidth{0pt} 
\tablecolumns{5}
\tablecaption{Summary of Observations}
\tablehead{\colhead{} 					&	 
           \colhead{UT Time} 				& 
	   \colhead{Time Resolution\tablenotemark{a}} 	&
	   \colhead{Number of} 				&
	   \colhead{} 					\\ 
	   \colhead{UT Date} 				& 
	   \colhead{(start of observations)} 		&
	   \colhead{(sec)} 				& 
	   \colhead{Exposures} 				& 
	   \colhead{Filter}				}
\startdata
2007 Dec 03 &01:55:30.0 &63.5 &285 &V\\
2007 Dec 04 &01:55:30.0 &63.5 &285 &B\\
2007 Dec 05 &01:38:30.0 &63.5 &315 &R\\
2007 Dec 14 &01:45:30.0 &63.5 &210 &I\\
2007 Dec 16 &02:04:30.0 &63.5 &203 &R\\
\enddata
\tablenotetext{a}{Mean time interval between exposures (integration time plus 
readout time)}
\end{deluxetable}

\clearpage

\begin{deluxetable}{lcr}
\tablenum{2}
\tablecaption{Times of Maximum Light}
\tablewidth{0pt}
\tablehead{	\colhead{HJD (peak)} 	& 
		\colhead{Cycle Number}		& 
		\colhead{$O-C$} 		\\
		\colhead{(2,450,000+)} 		& 
		\colhead{$(E)$} 		& 
		\colhead{($\times10^{-3}$~day)}	} 
\startdata
4437.616\dots &  0  &  $-2.306$  \\
4437.739\dots &  1  &  $-4.748$  \\
4438.628\dots &  8  &  $ 6.153$  \\
4438.747\dots &  9  &  $-0.289$  \\
4439.624\dots &  16 &  $-1.387$  \\
4439.754\dots &  17 &  $3.170$  \\
4448.658\dots &  88 &  $0.744$  \\
4450.663\dots & 104 &  $-1.337$  \\
\enddata
\label{time}
\end{deluxetable}

\clearpage

\begin{figure}
\epsscale{0.80}
\plotone{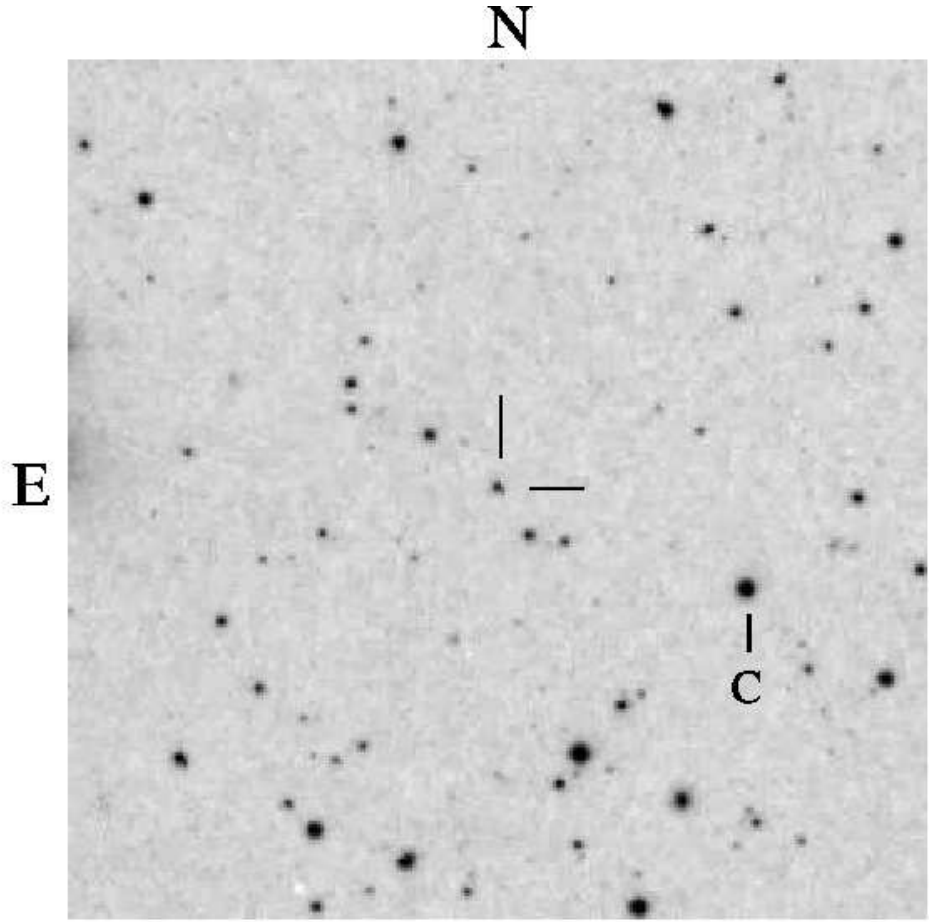}
\caption{The finding chart for Swift~J2319.4+2619. The coordinates of
          Swift~J2319.4+2619 are RA. = 23h~19m~30.43s,
          DECL. = +26$^{\circ}$~15$'$~19.1$^{''}$
          (equinox 2000.0) as given in Mukai et al. (2007).
          The comparison star used to calibrate our data is located
          $\sim2.5'$~W and $\sim1'$~S of Swift~J2319.4+2619, and is marked
          as star ``C". The scale is $\sim$8.5$'$ on a side.}
\end{figure}

\clearpage

\begin{figure}
\epsscale{0.80}
\plotone{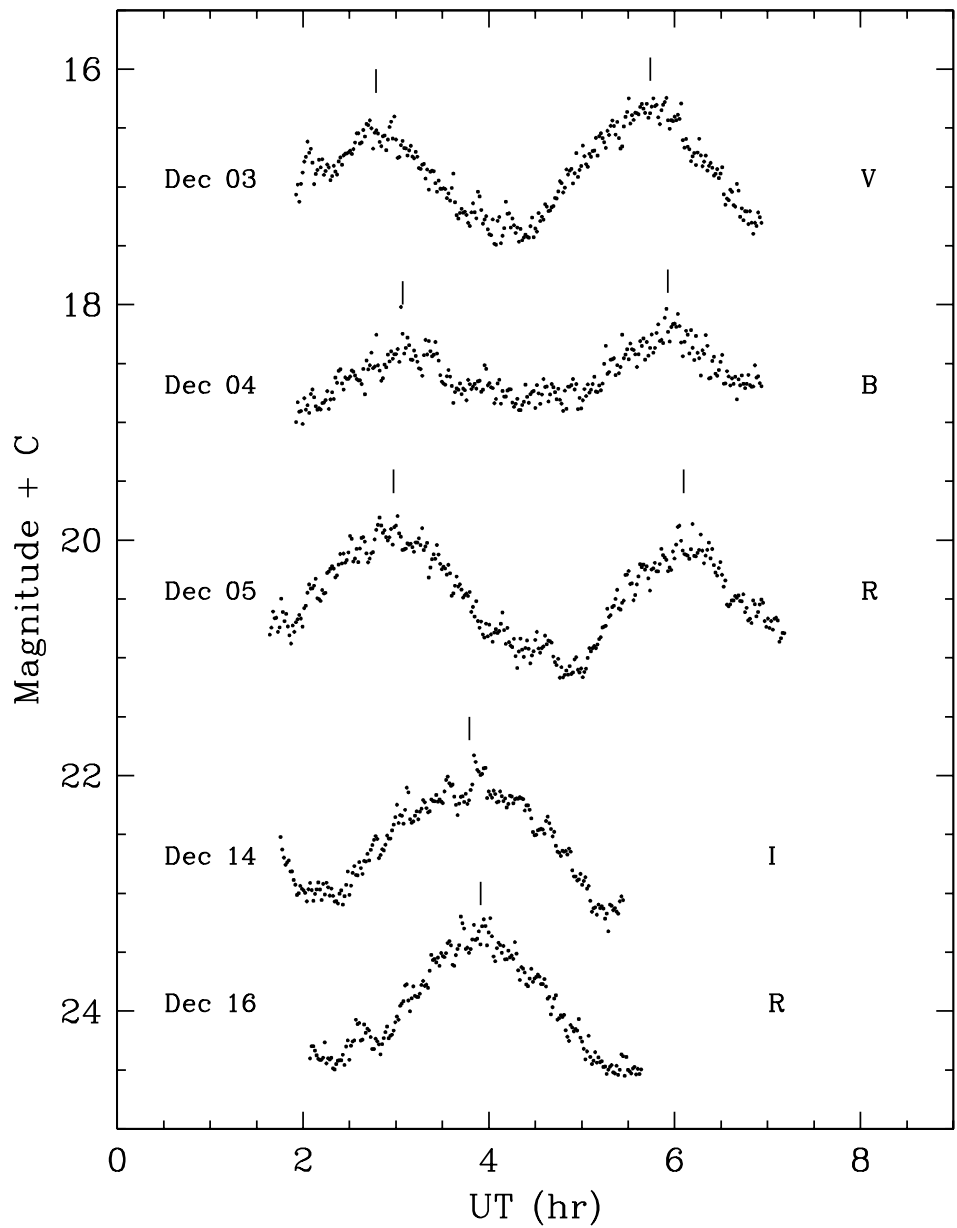}
\caption{The $B, V, R, {\rm and}~I$-band light curves of Swift~J2319.4+2619.
          The measured times of maximum are indicated by
          the vertical tick marks. Note that the amplitude of the modulation
          in $B$ is smaller than in the other colors. For clarity, the
          light curves have been offset by $C=1.5$, 4, 6.5, and 7.5~mag for
          the Dec~04 ($B$), Dec~05 ($R$), Dec~14 ($I$), and Dec~16 ($R$)
          light curves, respectively.
          }
\end{figure}

\clearpage

\begin{figure}
\epsscale{0.80}
\plotone{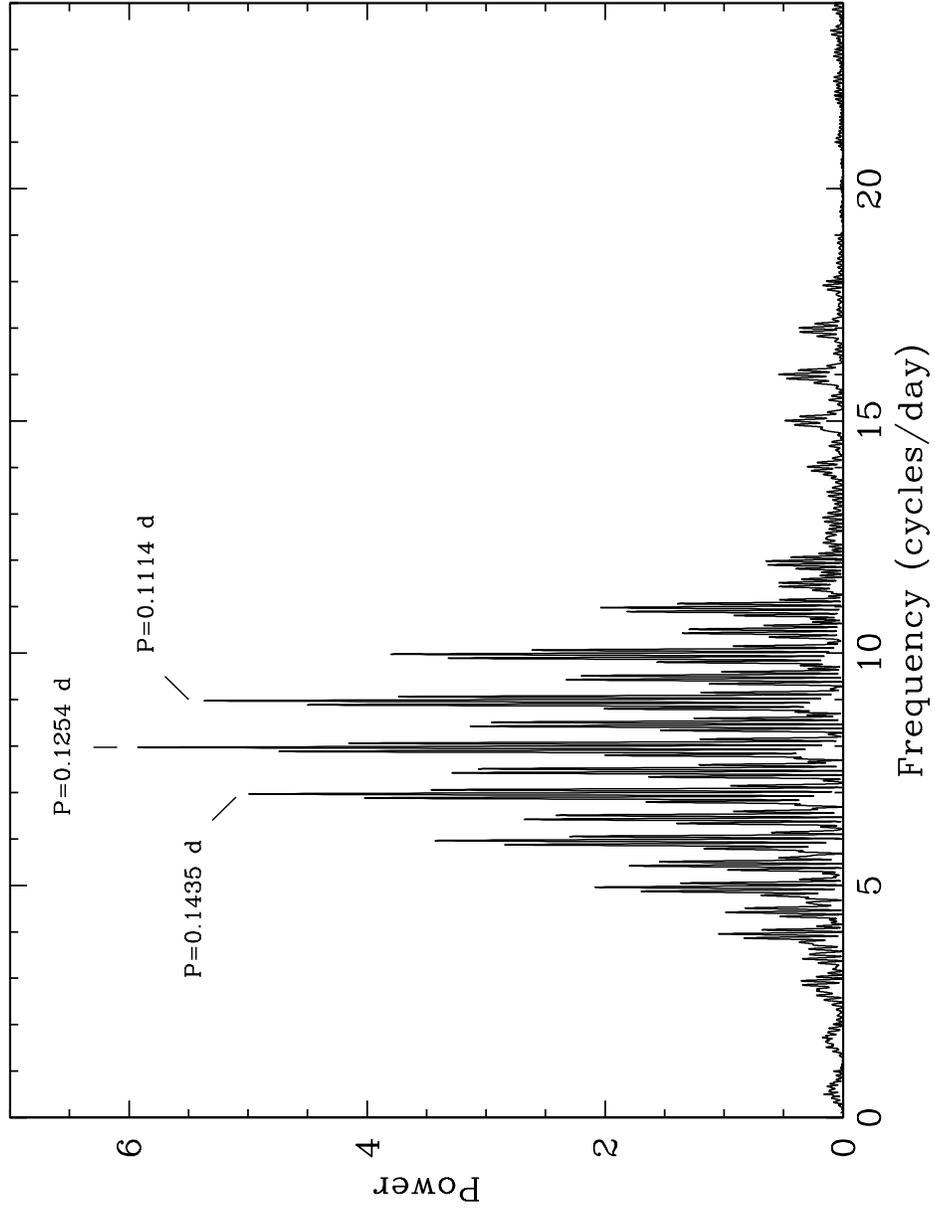}
\caption{The periodogram analysis of the complete
         set of light curve data.
         The highest peak corresponds to the
         favored period of $P=0.1254$~d, and is bracketed by one-day
         aliases of $P=0.1435$~d and $P=0.1114$~d.  
         }
\end{figure}

\clearpage

\begin{figure}
\epsscale{0.80}
\plotone{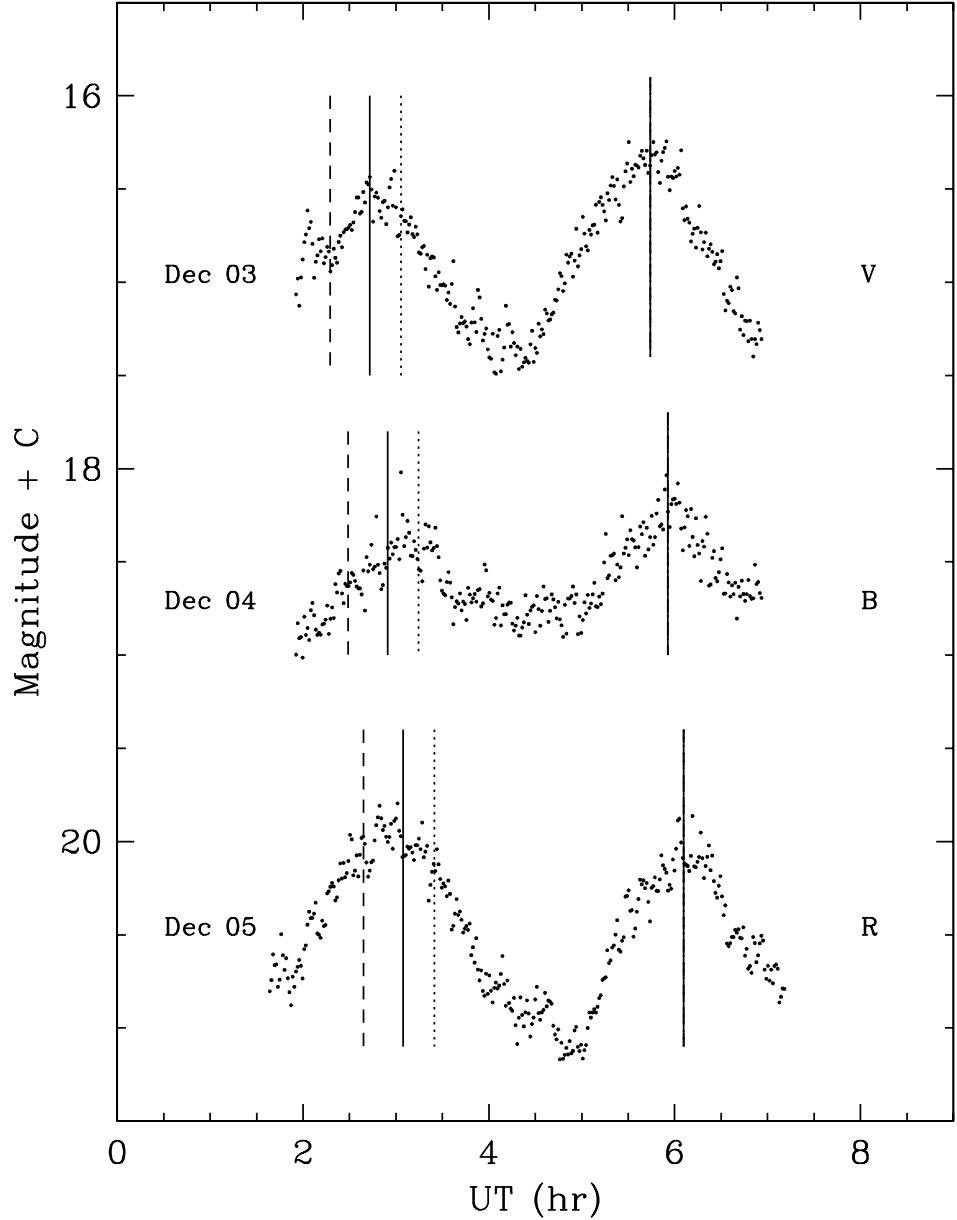}
\caption{The $B, V, R, {\rm and}~I$-band light curves of Swift~J2319.4+2619
          for the first three nights of data.
          For clarity, the light curves have been offset as in Fig. 2.
          The effect of choosing the one-day aliases periods in addition
          to the adopted period are shown
          (dashed line: $P=0.1435$~d, dotted line: $P=0.1114$~d, solid line:
          adopted period, $P=0.1254$~d).
          }
\end{figure}

\clearpage

\begin{figure}
\epsscale{0.80}
\plotone{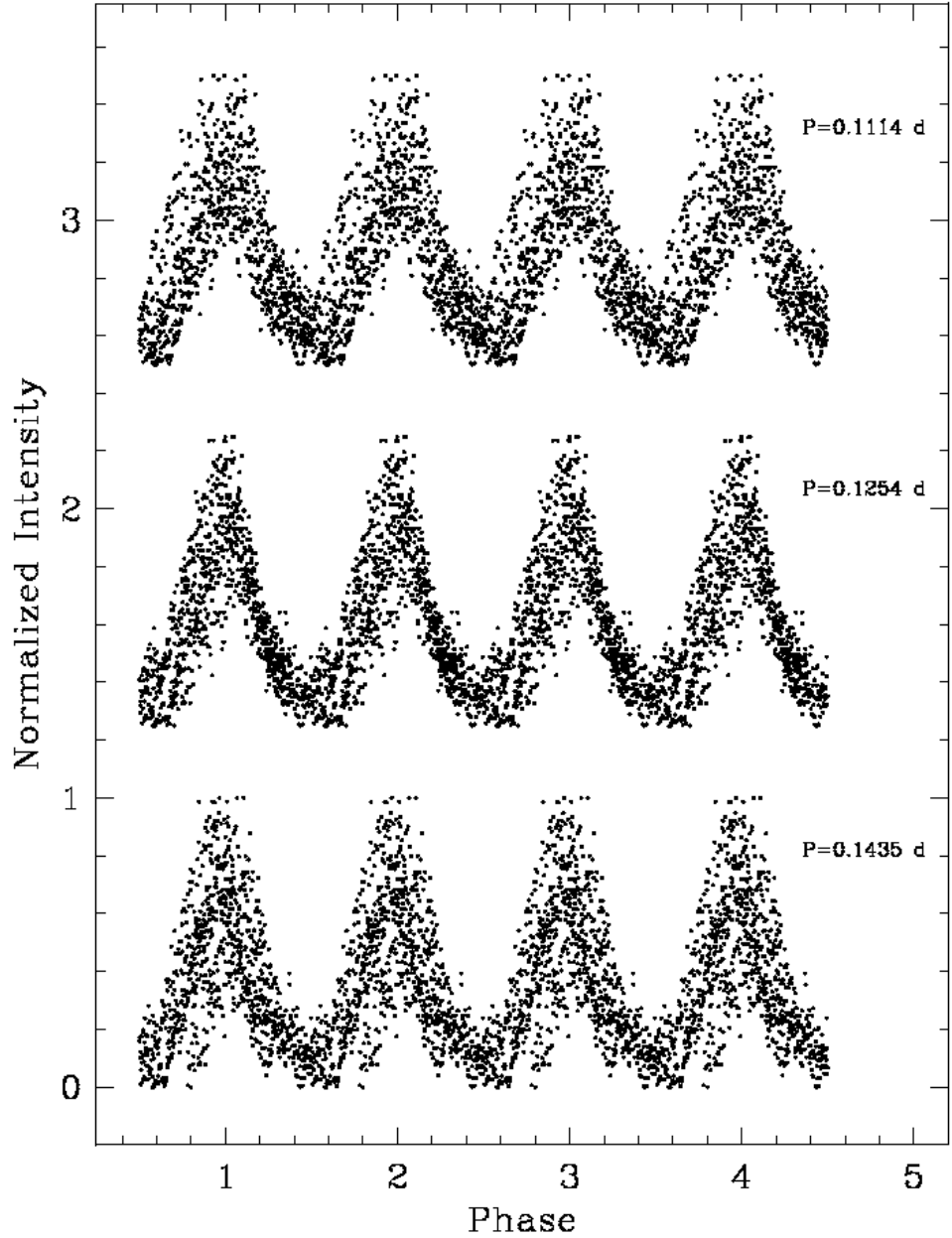}
\caption{The phased light curves for the first three nights of data
         based on the adopted period $P=0.1254$ (middle panel). For
         comparison, phased light curves based on the two one-day aliases
         periods, $P=0.1114$~d (upper curve), and $P=0.1435$~d
         (lower curve) are also shown. Note the tighter spread
         in the case of the adopted period.
         }
\end{figure}

\end{document}